\documentclass{epl}
\usepackage[dvips]{epsfig,geometry}
\usepackage{amsmath}

\pacs{61.30.-v} {Liquid crystals}
\pacs{42.79.Kr} {Display devices, liquid-crystal devices}
\title{Switching hydrodynamics in multi-domain, 
twisted nematic, liquid crystal devices}
\shorttitle{Switching in twisted nematic devices}
\author{D. Marenduzzo$^{1,2}$, E. Orlandini$^3$ and J.M. Yeomans$^1$}

\institute{$^1$ The Rudolf Peierls Centre for Theoretical
Physics, 1 Keble Road, Oxford OX1 3NP, England\\
$^2$ Mathematics Institute, University of Warwick,
Coventry CV4 7AL, England\\
$^3$ INFM, Dipartimento di Fisica, University of Padova,
Via Marzolo 8, 35131, Padova, Italy}

\begin{document}

\maketitle

\begin{abstract}
We study the switching dynamics in two-domain and four-domain
twisted nematic liquid crystal devices. The equilibrium configuration of these
devices involves the coexistence of regions characterised by different
handedness of the inherent director twist. At the boundaries between these
regions there are  typically disclinations lines. The dynamics of the
disclination lines controls the properties, and in particular the
switching speed, of the devices. We describe their motion using a
numerical solution of the Beris-Edwards equations of liquid crystal
hydrodynamics. Hence we are able to explain why a 
conventional
two-domain device switches off slowly and to propose a device design
which circumvents this problem. We also explain the 
patterns of disclination creation and annihilation that lead to
switching in the four-domain twisted nematic device.
\end{abstract}

\section{Introduction}

Twisted nematics (TN) are commonly employed 
in the construction of flat panel liquid crystal displays
\cite{degennes}.
In a TN device, the equilibrium configuration is one in which the
director field of the liquid crystal twists across the cell, 
normally because of
conflicting homogeneous anchoring at the boundaries.
Though traditional, single domain, TN devices can be built 
cheaply and easily, it is well known that 
they do not have ideal viewing angle properties~\cite{view1dtn}. 
To circumvent this, a number of possible solutions
have been suggested. One partially successful avenue has been 
the design and construction of multi-domain,
TN devices, in which regions of 
right-handed and left-handed twist 
alternate in the cell. This director structure is
imposed by using suitably patterned boundaries
\cite{atomic2dtn,kent,4dva,mdlcd1,mdlcd3,atn,stark}. 

While technological advances will likely make their production easier, 
there are fundamental problems associated with multi-domain TN devices 
that need to be investigated and understood theoretically. 
Most notably, because disclinations necessarily 
appear at the boundary between domains with different handedness, 
it is natural to expect that the disclination dynamics will
play a major role in determining the switching properties of the
devices. Understanding the disclination motion is vital in
suggesting better device designs.

Therefore in this letter we investigate how disclinations are created, move
and are destroyed as two- and four-domain TN devices are switched.
To do this we use a lattice Boltzmann algorithm to solve the 
three dimensional, Beris-Edwards equations of motion for 
nematic hydrodynamics. This formalism, where the equations are
written in term of a tensor order parameter, allows variation in the
magnitude of the order parameter and hence a natural description of
the disclination motion. This is a full hydrodynamic treatment which
includes the effect of back-flow.

We first consider a two-domain TN device, where stripes of left and 
right handed twist are separated by disclination lines.
We find that the application of an electric 
field pins the disclination line to one of the two surfaces. 
The switching on is slightly faster than for a single domain TN. 
The switching off is, however, considerably slower, 
as the driving force for the disclination to return to the centre of
the device is small. We suggest a novel device geometry aimed at
overcoming this problem.

We then present results for a four-domain device, which comprises a 
checkerboard of TN columns of different handedness. For this geometry
a surface pretilt or non-zero voltage are needed to sustain the
four-domain director pattern: otherwise the disclinations annihilate
leaving, effectively, either a single or two-domain TN device.
We find, in agreement with experiments, that for a given initial
four-domain state, very small fluctuations determine whether the
structure relaxes to a single or two domain TN and we describe
the dynamic pathways by which these states are reached.
We then follow the way in which disclinations are created, move and interact as
the field is switched on. These phenomena occur on the ms time scale.

\section{Equations of motion}

The equilibrium properties of the liquid crystal are described by a
Landau-de Gennes free energy density \cite{degennes,beris} (Greek letters
denote Cartesian components),
\begin{eqnarray}
f = \frac{A_0}{2}(1 - \frac {\gamma} {3}) Q_{\alpha \beta}^2 -
          \frac {A_0\gamma}{3} Q_{\alpha \beta}Q_{\beta
          \gamma}Q_{\gamma \alpha}
+ \frac {A_0\gamma}{4} (Q_{\alpha \beta}^2)^2 + 
\frac{K}{2}
\left(\partial_\alpha Q_{\beta \gamma}\right)^2
+\frac{\epsilon}{2}Q_{\alpha\beta}E_{\alpha}E_{\beta}
 \label{eqBulkFree}
\end{eqnarray}
where $K$ is the liquid crystal elastic constant,
$\epsilon>0$ for positive dielectric materials
(its value fixes the voltage scale, see captions to figures),
${\bf Q}$ is the tensorial order parameter 
(related to the director field $\hat n$ via
$Q_{\alpha\beta}=3/2 \langle n_{\alpha}n_{\beta}-1/3\delta_{\alpha\beta}
\rangle$ where $\langle\cdot\rangle$ denote a coarse grained average 
and $\delta$ is 
Kronecker delta \cite{degennes,beris}),
$\vec E$ is the electric field, $A_0$ is a constant and 
$\gamma$ controls the magnitude of order (quantified via the largest
eigenvalue of ${\bf Q}$, $q$). The
equation of motion for {\bf Q} is 
$D_t{\bf Q} = \Gamma {\bf H}$,
where $D_t$ is a suitable material derivative \cite{beris},
$\Gamma$ is a collective rotational diffusion constant
and  ${\bf H}= -{\delta f \over \delta {\bf Q}}+({\bf
    I}/3) Tr{\delta f \over \delta {\bf Q}}$.
The fluid velocity, $\vec u$, obeys the continuity equation and the
Navier-Stokes equation with a stress tensor appropriate to
liquid crystals (see Ref. \cite{beris} for details).
The director field configuration may give rise to a non-zero 
velocity via the stress tensor \cite{beris}: this 
effect is known as backflow.
To solve these equations we use a lattice Boltzmann algorithm, 
the details of which have been given in Refs. \cite{colin,proc}.
To locate disclination lines, we require that a point in the
simulation domain belongs to a disclination if the value of $q$ there
falls below $60\%$ of the average magnitude of order through the sample
(the thickness of the disclinations in Figs. 4 and 5 slightly depends on
the value of the threshold).\\
 
\section{Two-domain twisted nematic device}

\begin{figure}
\centerline{\psfig{figure=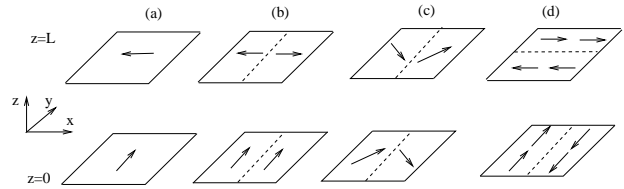,width=3.2in}}
\caption{The geometry considered in the
text, together with the surface patterning leading to a (a) single
domain (b,c) 
two-domain and (d) four-domain TN device. }
\label{setup}
\end{figure}

The boundary patterning necessary to stabilise a TN device is shown in
Fig. \ref{setup}a, together with the definition of the 
co-ordinate axes we shall use (periodic boundaries are taken along
$x$ and $y$).
We first consider a two-domain twisted nematic device (2DTN), in
which left-handed and right-handed twisted nematic stripes are separated by 
disclinations. The boundary patterning necessary to stabilise the 2DTN 
is shown in Fig. \ref{setup}b. The handedness of different sections of the 
device is imposed by rubbing the surfaces such that the director field has 
a $\pm2^{\circ}$ pretilt (ie angle to the $xy$-plane).

Without an electric field, there is 
a twist disclination line parallel to the $y$ axis 
at the centre of the cell between domains of
opposite handedness. In equilibrium no escape into the
third dimension is found at the disclination core
in the one elastic constant approximation,
if there is no pretilt. If we relax the one elastic 
constant approximation, e.g. by
adding the term $K'\left(\partial{_\beta}Q_{\alpha\beta}\right)^2$
to $f$ in Eq. \ref{eqBulkFree}
where $K'$ is another elastic constant \cite{degennes}, 
we find $n_z \ne 0$ at the disclination core ($n_z=0.08$ for
$K'/K=2$ with other parameters as in Fig. 2).

\begin{figure}
\begin{center}
{\psfig{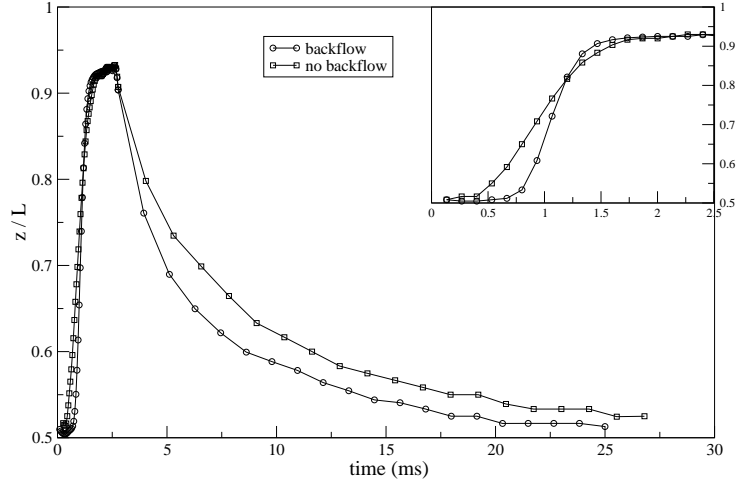}} 
\end{center}
\caption{Switching dynamics in a 2DTN device. The plot 
shows the time evolution of the $z$ position of the disclination, for the
two cases with and without backflow.
Parameters are $K=25$ pN, $L=1.5$ $\mu$m, $\gamma=3$,
the average magnitude of order is $q=0.5$,
the applied voltage is $2.4$ $V_c$, where $V_c$ is
the Freedericksz threshold \cite {degennes} 
(in a Freedericksz cell with the same parameters), 
and there is a pretilt of $2^o$ at 
the surfaces. We used
$A_0/(KL^2)=7$ $10^{-4}$.
The rotational and isotropic viscosities (see Ref.
\cite{proc}) are $\gamma_1=1$ and $\eta=0.03$ Poise.
The field was switched on at $t=0$ and off at $t=2.7$ ms. The
inset shows the switching on dynamics.} 
\label{switching}
\end{figure}

Our aim is to understand how the disclination 
affects the switching dynamics.
Fig. \ref{switching}a shows the
position of the disclination as a function of time 
during a cycle of turning the field on, and then off.
When the field is switched on the disclination moves quickly
to the surface at $z=L$. It is advantageous to accomodate the
disclination here because of the particular geometry of the 
surface pretilt.
When the field is switched off, the disclination line
moves slowly back to the centre of the device.
This is a much slower process because, in zero field
the free energy of the state with the
disclination at the surface of the device is only slightly 
higher than that with it at the centre.
Therefore the driving force is small and the disclination
will be easily pinned en route by any
impurities.

We ran simulations with purely relaxational dynamics 
and compared them to the physical case where movements
of the director can induce flow fields in the device.
The results are compared in Fig. \ref{switching}.
The effect of backflow is to delay the onset of switching when the
field is applied but to enhance the speed of the
disclination once it starts to move. Backflow also 
increases the speed with which the disclination returns to the
centre of the device when the field is switched off. 

While the switching on is slightly
faster for the two-domain device than for
a conventional single-domain TN device with the same conditions, 
the switching off
is substantially slower
as the disclination takes a long time to move back to the centre of the cell. 
This is obviously deleterious to its operation as a fast display. To
overcome this problem we propose a two-domain device where the domain
structure is stabilised by a surface $\pi/2$ pre-twist (see Fig. 1c).
The disclinations now remain at the surface in both the on and off
states.

Fig. 3 compares switching
in two-domain 
devices with pre-tilt (as in Fig. 1b) and with a 
$\pi/2$ surface pre-twist (as in Fig. 1c), with identical
parameters. The value of the director in the $xy$ plane and along the
$z$-axis are shown as a function of time. Both devices would lead to
a comparable viewing angle improvement over the single domain
cell. The pre-twist device
switches on as quickly as the pre-tilt device. 
However it avoids the slow switching off of the pre-tilt device.

\begin{figure}

\begin{center}
{\psfig{figure=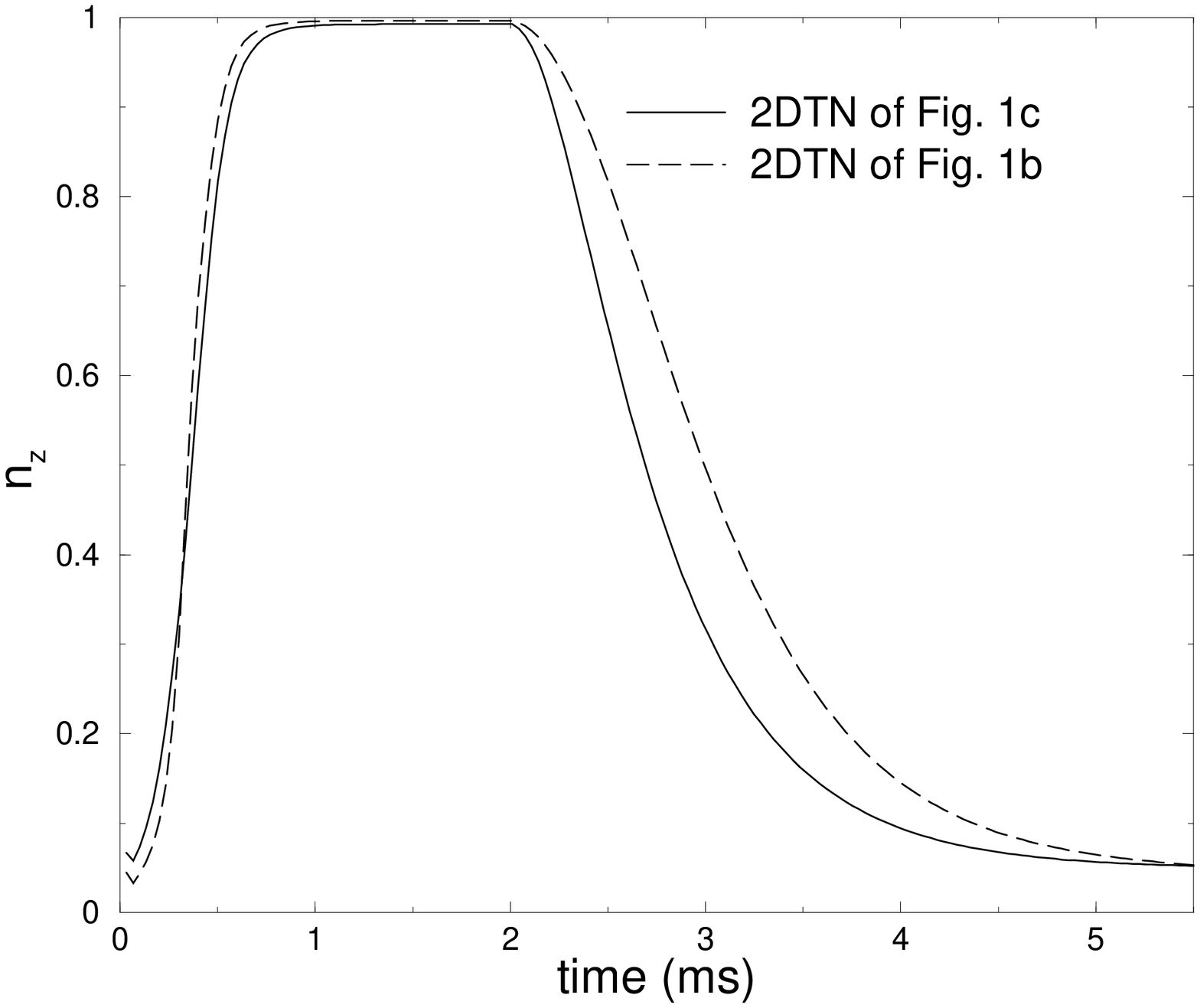,angle=270,width=2.3in}
\psfig{figure=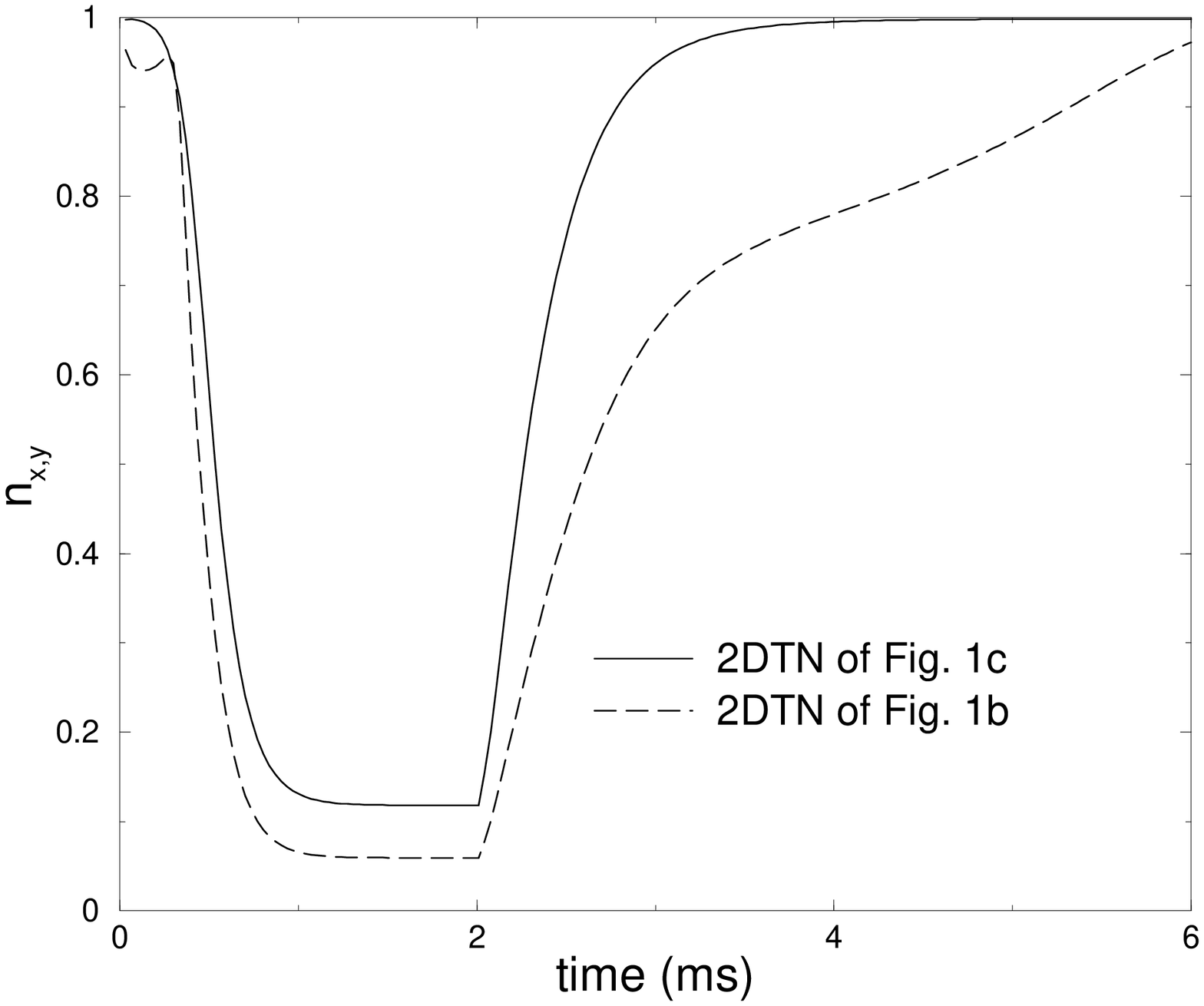,angle=270,width=2.3in} }
\end{center}
\caption{Comparison between switching dynamics in a 
two-domain twisted nematic device as in Fig. 1b, and in another one 
as in Fig. 1c, in which the disclination 
is at the surface for all times. 
The plot shows the time evolution of the $z$ component of the
director field along $z$ ($n_z$, left) and of its larger component
on the $xy$ plane ($n_{x,y}$, right). Backflow is not considered.
Here $L=0.75$ $\mu$m, $V=1.92$ $V_c$. Other parameters are as in Fig. 2.}
\label{switching_surface}
\end{figure}

\section{Four-domain twisted nematic device}

We now consider how disclination dynamics influences the switching 
of a four-domain TN (4DTN) liquid crystal device. In
such a device the target is to obtain a configuration in which
columns of left- and right-handed twist are
organised in a checkerboard pattern. The rubbing directions
we consider are shown in Fig. \ref{setup}d: this mimicks the
experimental set-up chosen in Refs. \cite{kent}.
Note that, unlike the 2DTN device, the two surfaces of the
4DTN device can carry the same patterning.
A related theoretical study,
within the Ericksen-Leslie theory, 
has been 
proposed in Ref.~\cite{stark}.

In the 4DTN device, the pre-tilt angles 
on or near the surfaces are crucial to
the stabilization of the director field configuration.
This can be understood within a simple model~\cite{kent}. 
Creating the disclinations increases the free energy
and therefore the surface must be rubbed 
with a non-zero pretilt, $t$, so that removing
the disclinations will lead to a greater free energy
penalty:
if there is no pretilt the 4DTN display is never stable but unwinds. 
There is a finite
critical threshold for the pretilt, $t_{\rm min}$, which can stabilise the 
structure at zero voltage. This depends on the geometry of the
device~\cite{kent}. For a pretilt
$t$ smaller than $t_{\rm min}$, the four-domain
structure is not stable at
low voltages. Experiments~\cite{kent} and theory~\cite{kent,stark} 
show however that, even for $t<t_{\rm min}$, 
the 4DTN device is stable for voltages above a 
certain threshold $V_{\rm min}(t)$.

\begin{figure}
\begin{center}
\begin{tabular}{ccc}
(1a) 0.7 ms & (1b) 2.0 ms & (1c) 3.4 ms \\
\psfig{figure=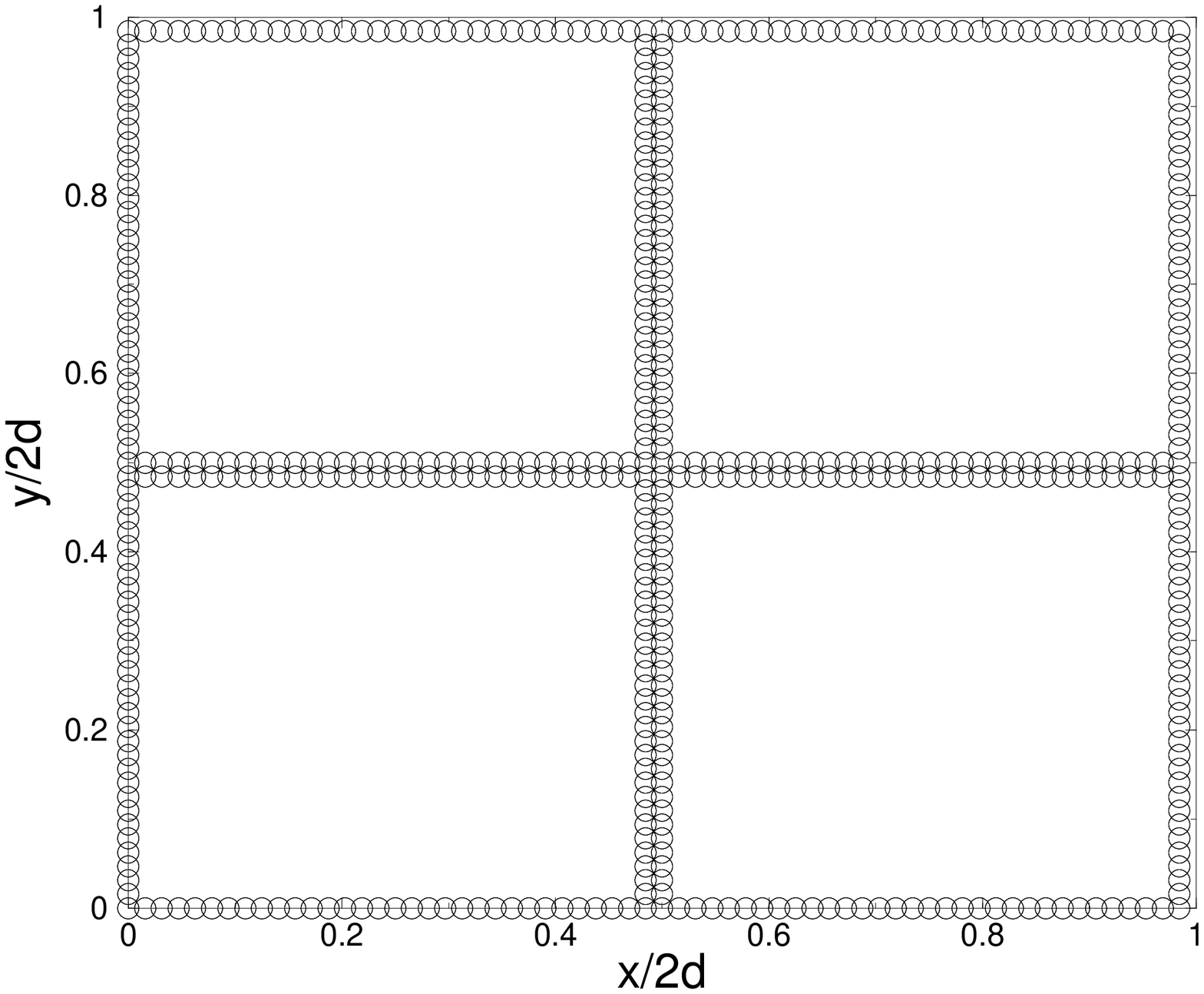,angle=270,width=3.6cm} &
\psfig{figure=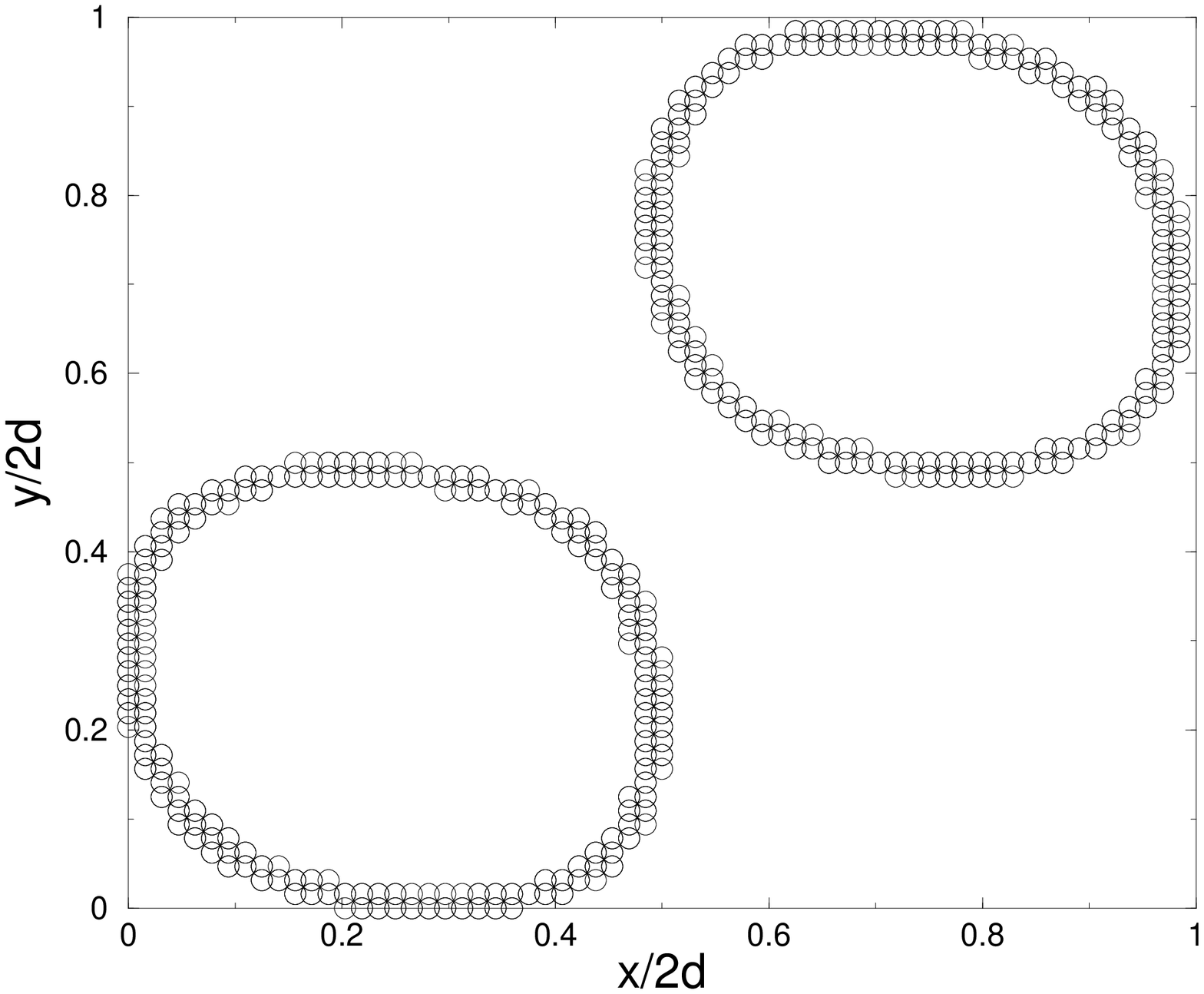,angle=270,width=3.6cm} &
\psfig{figure=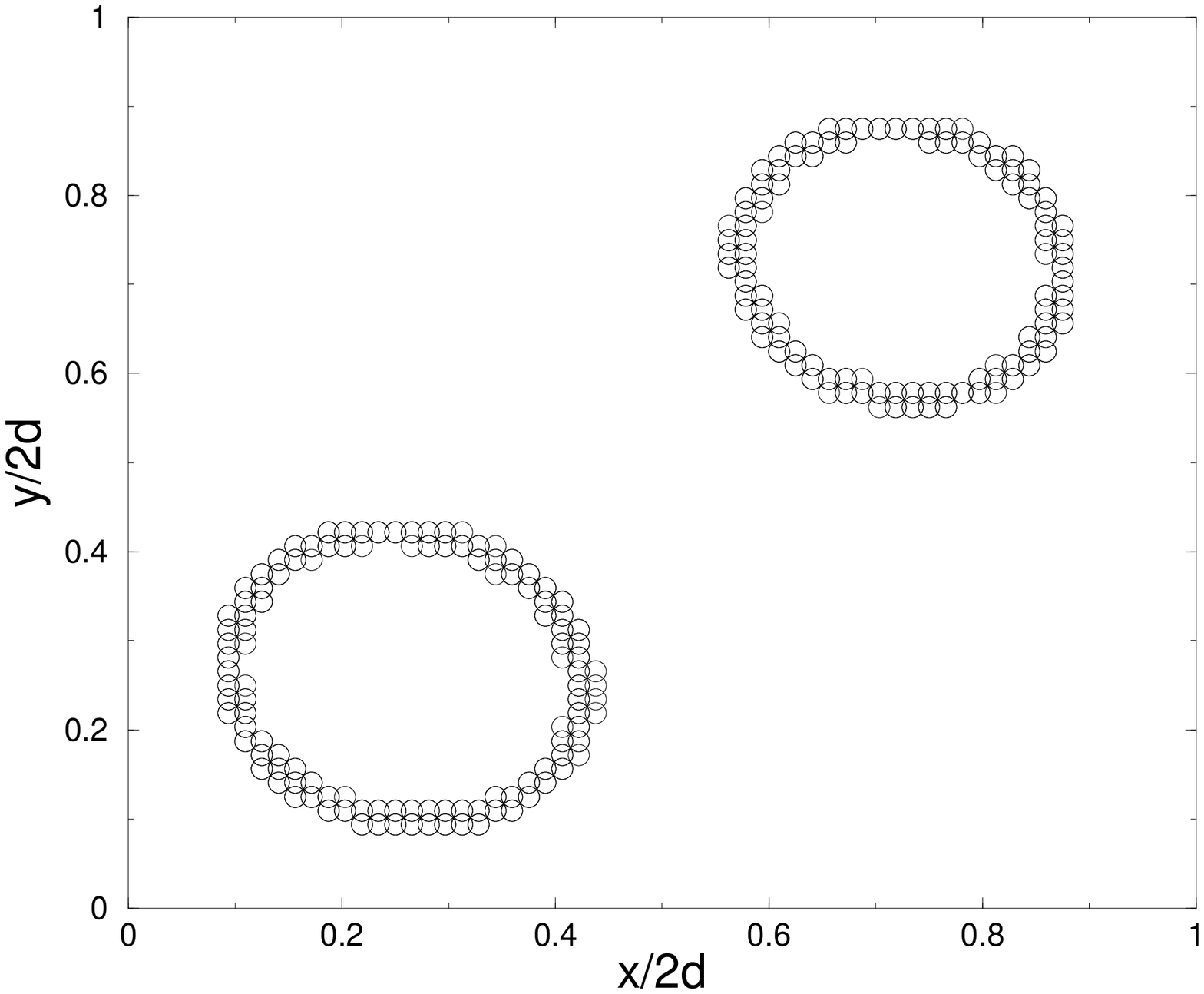,angle=270,width=3.6cm} \\
(2a) 0.7 ms & (2b) 2.0 ms & (2c) 6.7 ms \\
\psfig{figure=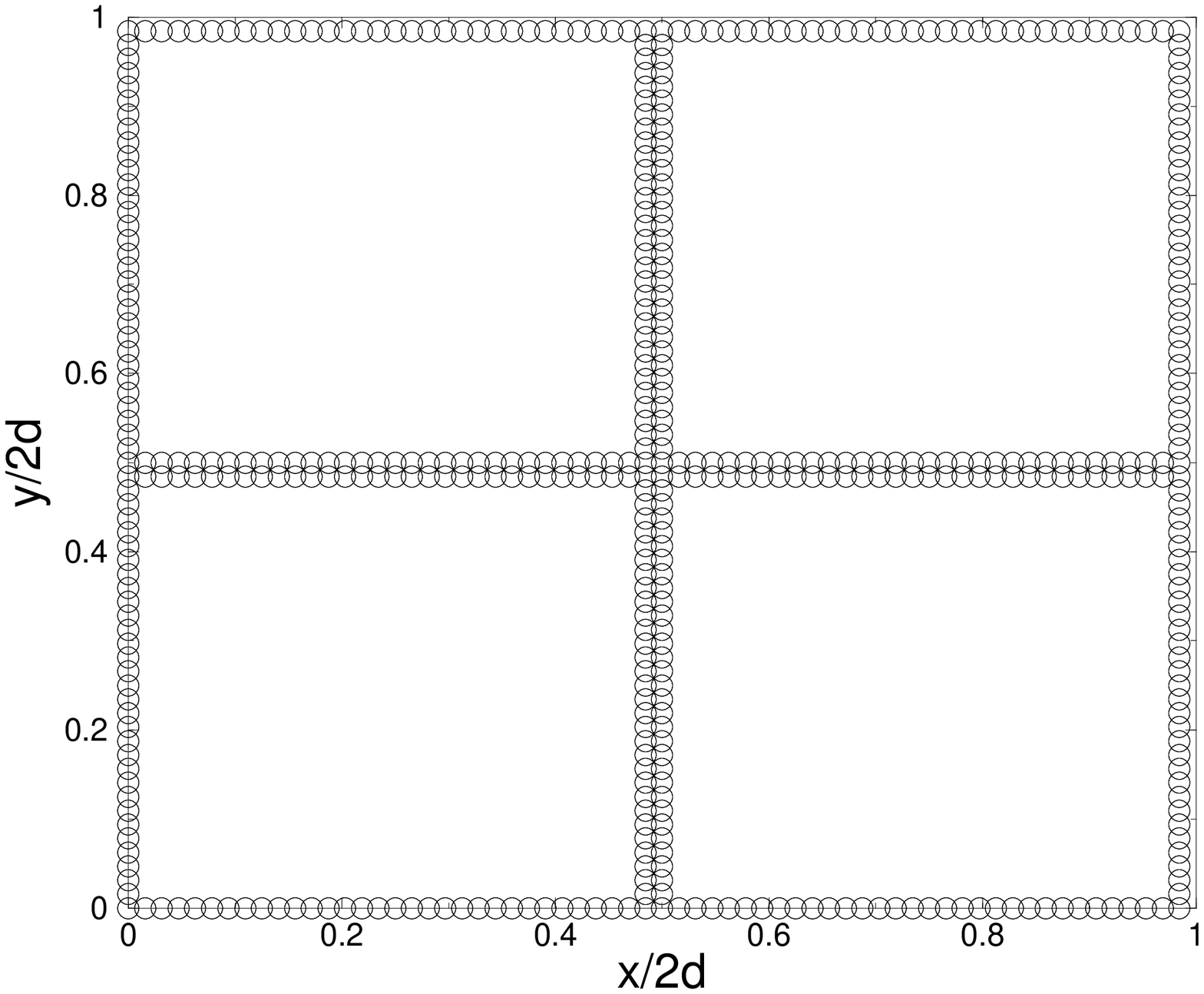,angle=270,width=3.6cm} &
\psfig{figure=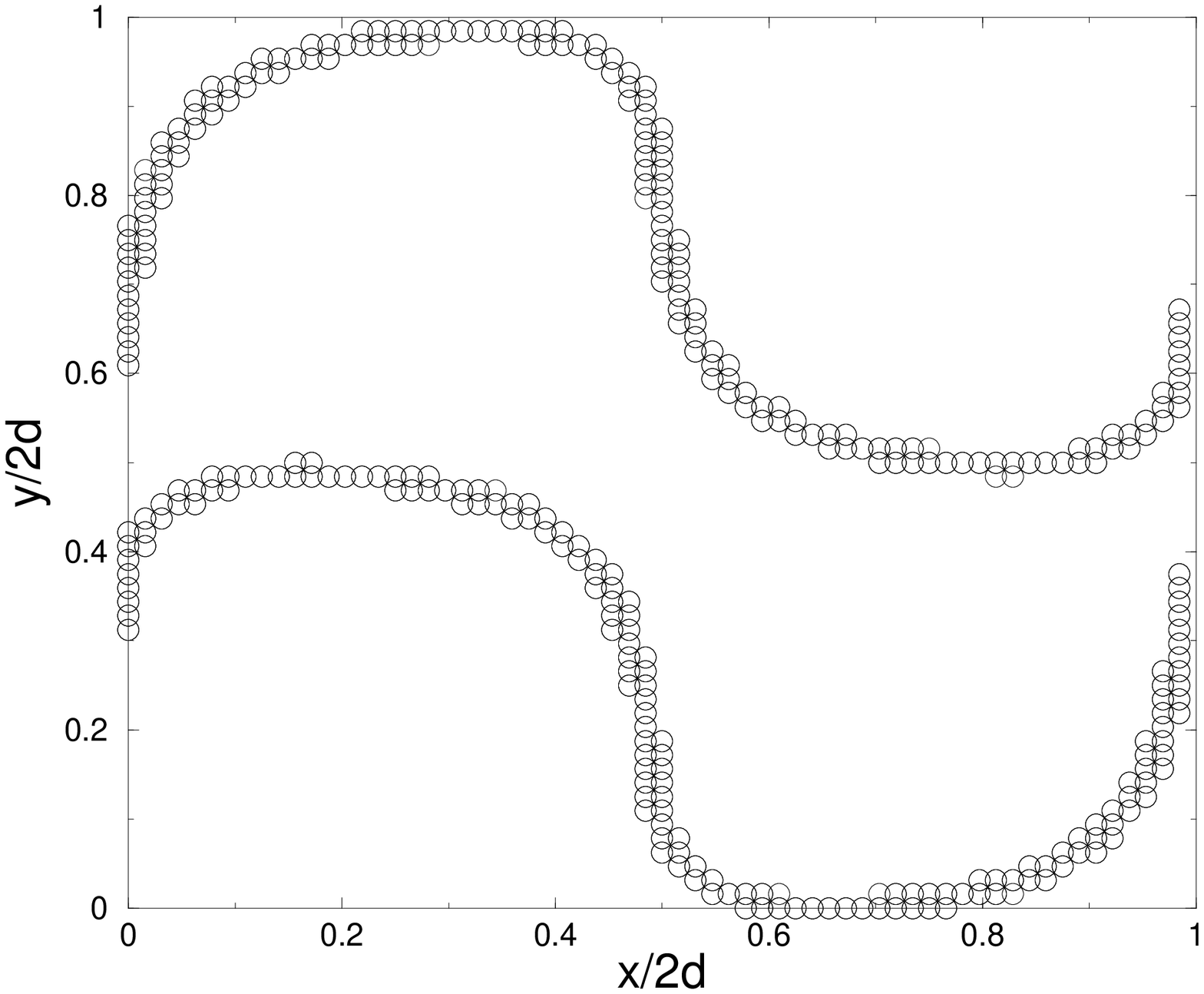,angle=270,width=3.6cm} &
\psfig{figure=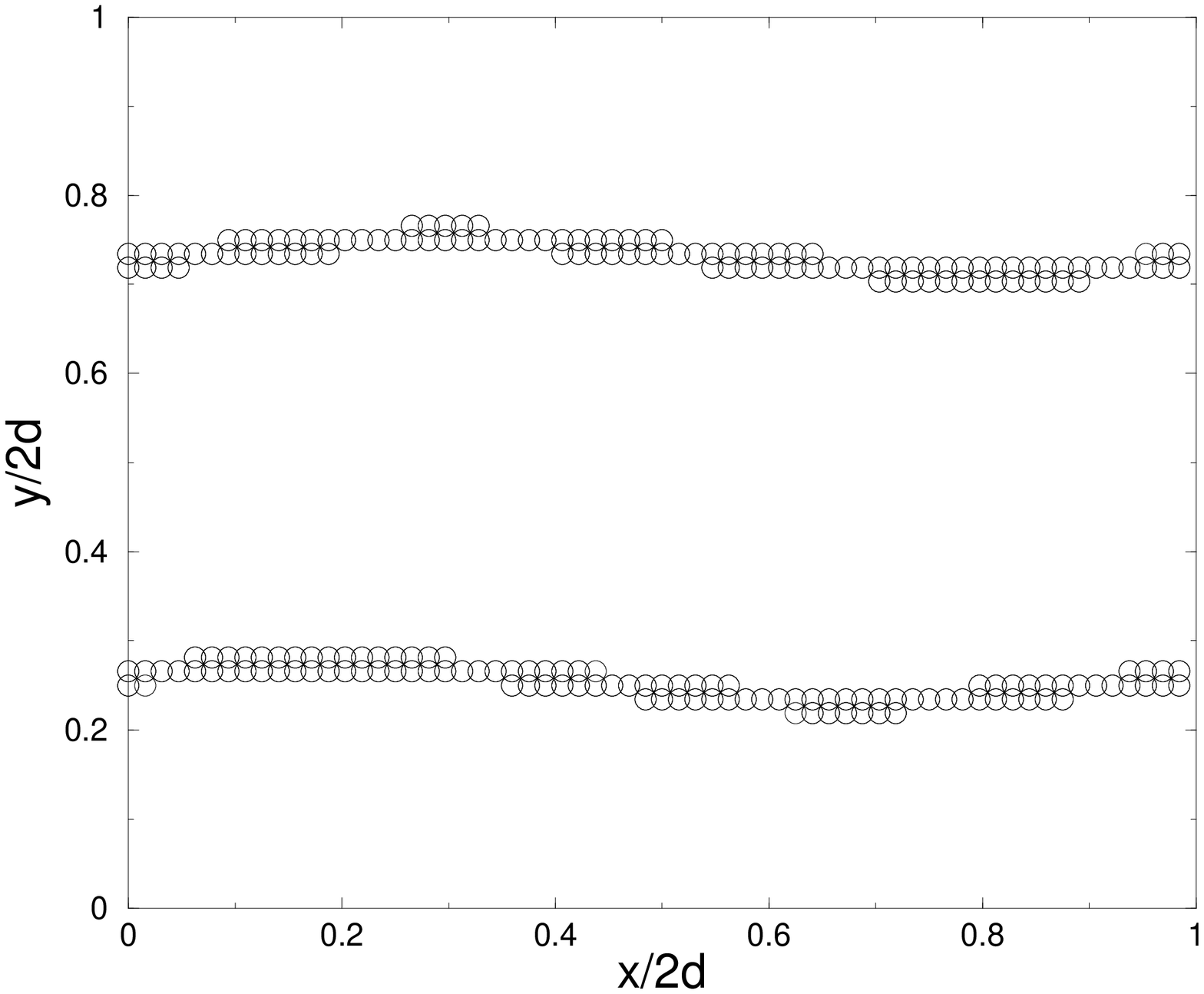,angle=270,width=3.6cm} \\
\end{tabular}
\caption{Two annihilation pathways in a 4DTN
device with a pretilt $t_0=10^{\circ}$ for 
$V<V_{\rm min}(t_0)$ so that the
disclinations  annihilate. Only the midplane of the device ($z=L/2$) 
is shown as the disclinations are
confined there throughout the simulation.
These pathways correspond to those
experimentally observed in Ref. [5].
Parameters are: $K=25$ pN, $L=1$ $\mu$m, $\gamma=3$ and
$\gamma_1=1$ Poise. The average magnitude of order is $0.5$.
We used $A_0/(KL^2)=0.025$.
Applied voltages are: 
$0.75$ $V_c$ (top) and $0.15$ $V_c$ (bottom), 
$V_c$ being the Freedericksz threshold (as in Fig. 2).
The domain dimension is $2d\times 2d = 
3.2$ $\mu$m $\times$ $3.2$ $\mu$m.}
\label{4Dstabilization}
\end{center}
\end{figure}

In our numerical simulations, we consider a pretilt $t_0=10^{\circ}$ which 
is less than $t_{\rm min}$. For voltages smaller than 
$V_{\rm min}(t_0)\sim 0.9$ $V_c$ ($V_c$ is the Freedericksz threshold)
the four domain structure is not stable and reduces 
to what is effectively a single domain
or a two-domain twisted configuration. The single domain has lower free 
energy, but the two domain structure is also a local minimum. 
Typical pathways followed by the disclination lines as the 
4DTN structure unwinds, are shown in 
Fig. 4. It is to be stressed that the
different pathways follow from {\em identical}
initial conditions: which pathway is followed
depends solely on (numerical) noise. 
These patterns 
resemble those observed in the experiments in
\cite{kent}. 

Above $V_{\rm min}(t_0)$, the 4DTN device  is stable and
there are well-defined  columns of right and left-handed twist. Our
simulations show however that at these voltages the disclinations  no
longer lie at the centre of the cell but pin to the top or bottom
surfaces between the  regions of different pretilt in such a way that
they do not cross.  Near the centre of the cell there is a large
component of the director field along the $z$ direction.

\begin{figure}
\centerline{\psfig{figure=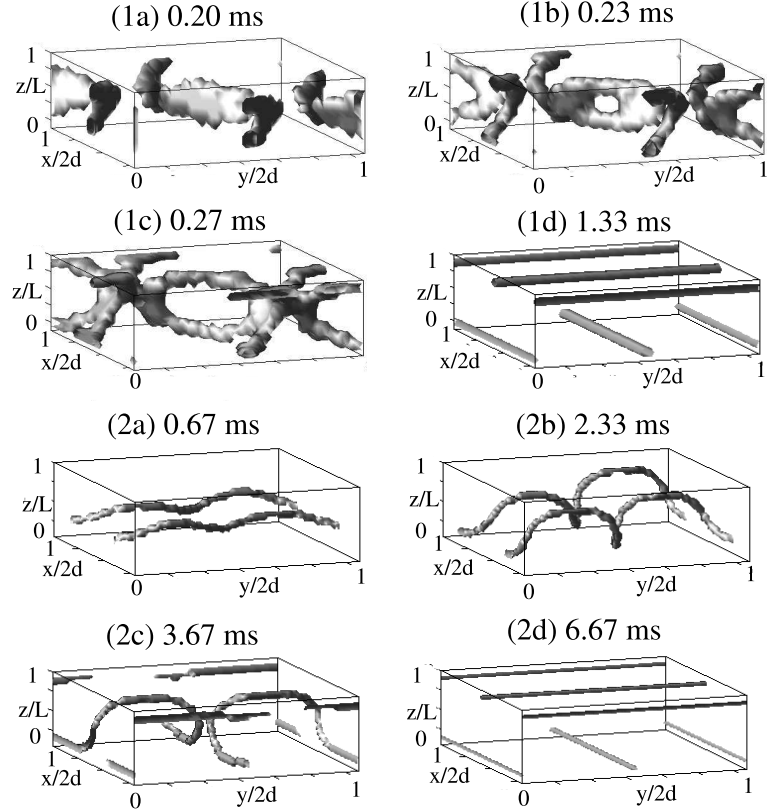,width=4.in}}
\caption{Evolution of the disclination configuration as a 4DTN
device is switched on from
the stable disclination free configuration (frames 1a-1d) and 
 from the metastable configuration shown in Fig. 4 (second row) 
(frames 2a-2d). 
The voltage at $t=0$ is smaller than $V_{\rm min}(t_0)$. It
is changed to a value larger than $V_{\rm min}(t_0)$
after $6.7$ or $13.4$ ms for the pathways 1 and 2 respectively.
The voltages are 0.60 $V_c$ and $1.79$ $V_c$
for frames (1a-1d) and $0.15$ $V_c$ and $1.34$ $V_c$ for frames (2a-2d).
The time, measured from the moment the voltage has been increased,
is shown above each frame. The domain dimension is $1.6$ $\mu$m $\times$ 
$1.6$ $\mu$m for the first pathway and
$3.2$ $\mu$m $\times$ $3.2$ $\mu$m for the second. $L=0.5$ (and 
$A_0/(KL^2)=0.025$) and $1$ $\mu$m (and 
$A_0/(KL^2)=0.1$)
for the top and bottom pathways, while 
other parameters are as in Fig. 4.}
\label{4d_switching}
\end{figure}

We now describe the dynamics of the switching from the
off state in which the voltage is below the critical value
(and $t<t_{\rm min}$ so the four-domain structure has collapsed), 
to the on state in which the disclinations are pinned at 
the device boundaries.
In Fig.~\ref{4d_switching} we show the dynamical switching 
pathways starting from a stable disclination-free 
configuration (frames 1a-1d), and 
from the metastable configuration shown in Fig.
\ref{4Dstabilization} (frames 2a-2d).

In the first case, disclination loops 
are created in the middle of the sample. These 
loops then grow (Fig.~\ref{4d_switching}, 
frame (1b)) until they
touch the boundaries and then approach each other
sufficiently closely that they can interact
(frame 1c). They annihilate at the centre of 
the device and merge at the boundaries to form the pinned 
disclination pattern characterising the large voltage stable state.
A similar switching mechanism was postulated in \cite{stark}.

In the second case shown in Fig.~\ref{4d_switching},
frames 2a-2d, two disclinations
are present at the beginning of the simulation in a metastable starting state.
This configuration is likely to be present in the experiments in some 
regions of the sample~\cite{kent}. After the voltage is increased above
the critical threshold, the disclinations bend (frame 2a) and
different sections of the disclination lines 
start moving to the two surfaces. 
During this motion, the disclinations acquire a non-zero twist 
(frames 2b-2c).  
The disclinations distort when they touch the boundary and move towards
the stable configuration, which is 
again reached by merging the disclinations at the boundaries 
and annihilating those at the centre of the cell. \\

\section{Discussions and conclusions}

In conclusion, we have solved numerically
the hydrodynamic equations of motion to
show that in the two- and
four- domain TN devices disclination dynamics is crucial
to the switching. We hope that understanding the disclination motion
will prove useful in improving the design
of multi-domain TN devices. For example, in the two-dimensional TN device,
we found that the switching off is slowed considerably by having to wait 
for the disclination to return to the centre of the cell. We showed that, 
if the surfaces are
patterned with an alternating large ``pre-twist'' 
(in a manner similar to that
described for related multi-domain devices in \cite{mdlcd3}),
the disclinations sit at the surface of the device even at zero voltage.
Hence switching occurs without any need to
rearrange the disclination pattern, thus circumventing the long time
needed for switching off whilst preserving the viewing angle
advantages of the two-domain device.
This and the other predictions made here could be tested in devices of 
suitable 
dimensions, by deducing the disclination line position from the time dependent
optical properties of the sample.

\end{document}